\documentclass{ws-jtcc}
\usepackage{graphicx}
\begin{document}

\markboth{K. Doll, T. Jacob}
{QM/MM description of periodic systems}

%
\catchline{}{}{}{}{}
%

\title{QM/MM DESCRIPTION OF PERIODIC SYSTEMS}

\author{K. DOLL}

\address{Institut f\"ur Elektrochemie, Universit\"at Ulm,
Albert Einstein-Allee 47, D-89081 Ulm, Germany
klaus.doll@uni-ulm.de}

\author{T. JACOB}

\address{Institut f\"ur Elektrochemie, Universit\"at Ulm,
Albert Einstein-Allee 47, D-89081 Ulm, Germany\\
Helmholtz-Institut Ulm (HIU), D-89069 Ulm, Germany}

\maketitle

\begin{history}
\end{history}

\begin{abstract}
A QM/MM implementation for periodic systems is reported. This
is done for the case of molecules and for systems with two and
three-dimensional
periodicity, which is suitable to model electrolytes in contact with
electrodes. Tests on different water-containing systems, ranging from the
water dimer up to liquid water
indicate the correctness of the scheme. 
Furthermore, molecular dynamics simulations are performed, as a possible
direction to study realistic systems.
\end{abstract}


\section{Introduction}

The theoretical description of extended systems has become more and more
important in the course of treating realistic systems. 
Presently, energy conversion has become
an important
topic, and there has been an upsurge of research in electrochemistry.

The targeted systems are rather complex, {\it e.g.} the simulation of
electrolytes in contact with electrodes. This can often not be done with
density functional theory, due to the system size and
corresponding large numerical effort.
On the other hand, empirical models may
not always work well, and besides, they depend on the choice
of parameters, such as force fields. The combination of quantum mechanical (QM)
and molecular mechanics (MM) methods (QM/MM) is therefore tempting in this area.
Therefore, the present article aims at a description of an implementation
of such a QM/MM scheme for molecular and periodic systems.

\section{Method}
\label{methodsection}

The main intention is to establish a QM/MM scheme for molecules {\it and}
periodic systems. These schemes have become popular over the past years
especially for the modeling of biological systems, see
 {\it e.g.}
 \cite{Maseras1995,Svensson1996,Eichler1996,Bakowies1996,Sherwood2003}.
Also, periodic systems are now investigated on this level,
{\it e.g.} \cite{Sauer2000,Yarne2001,Laino2006,Sanz2011}.
The systems to be targeted with the present approach
are in the area of electrochemistry, {\it e.g.}
electrolytes, in contact with metal surfaces. 
Motivated by the aim to develop a theoretical scheme to treat electrochemical
interfaces self-consistently, we formulated a QM/MM approach which in the
first stage is making use of the so called subtractive scheme to evaluate
the system's energy
(see, {\it e.g.} \cite{Sherwood2003}). In this scheme,
the total energy and forces are obtained as

\begin{eqnarray} & 
E_{\mbox {QM/MM}}^{\mbox {whole system}} = &
E_{\mbox {MM}}^{\mbox {whole system}}+ 
 E_{\mbox {QM}}^{\mbox {QM region}}
-E_{\mbox {MM}}^{\mbox {QM region}}
\label{energyequation}
\end{eqnarray}

\begin{eqnarray} &
\vec F_{\mbox {QM/MM}}^{\mbox {whole system}} = &
\vec F_{\mbox{MM}}^{\mbox {whole system}}+
 \vec F_{\mbox{QM}}^{\mbox {QM region}}
-\vec F_{\mbox{MM}}^{\mbox {QM region}}
\label{gradientequation}
\end{eqnarray}

Compared to molecules, the periodicity has to be taken into account. This is
in the present approach
achieved by summing the interactions with the help of the Ewald method in the
case of periodicitiy in three dimensions, and Parry's potential 
\cite{Parry} for
the case of periodicity in two dimensions. The MM interaction is thus summed
for the whole system, and subsequently corrected due to the contributions
from the QM region: the energy of the QM region and its periodic replicas
are again summed with the Ewald or Parry method, on the MM and the QM level
separately. Subsequently, the data obtained is inserted in equations
\ref{energyequation} and \ref{gradientequation}. This QM/MM level
corresponds to a mechanical embedding, where the interactions between the
QM and MM region are treated on the MM level. The QM and MM calculations
for the QM region are performed without taking the atoms into account
which are not in the QM region. However, the total energy in
equation \ref{energyequation} and the forces in equation
\ref{gradientequation} now include effects of the QM region.
An extension of mechanical embedding leads to
electrostatic embedding
where the QM calculations include the effects of all the other atoms by
e.g. representing these atoms by point charges. The present implementation
corresponds thus to a periodic extension of a mechanical embedding. For
a comparison of the embedding schemes, see e.g. \cite{LinTruhlar2007}.

\section{Implementation}

The present scheme requires a QM and an MM code, and a connection between
both. For the MM region GULP was used \cite{Gale1997}, and
for the QM region CRYSTAL \cite{Manual09}. Energies and gradients
are obtained from these codes and manipulated according to equations
(\ref{energyequation}) and (\ref{gradientequation}). Besides these codes,
it is necessary to have 
a routine which works with the manipulated energies and forces: here,
a modified version of GULP is used. 
Finally, an interface ({\it i.e.} wrapper)
to perform the communication between the three
aforementioned codes is required, 
for purposes such as {\it e.g.} extracting the energy and gradients,
or setting up new input files during an optimisation ({\it e.g.} GULP and
CRYSTAL
inputs for the QM region are generated automatically during an optimisation).

This strategy has the advantage that features which are
available within GULP can be used for QM/MM calculations, 
such as the geometry optimisation
or molecular dynamics. However, one has to keep in mind that in
the present implementation
only the MM Hessian is available.

\section{Computational parameters}
\label{technicaldetails}
The target of the present work is mainly to establish the methodology. 
A relatively simple potential
is used
\cite{ToukanRahman1985}, which employs intermolecular
Coulomb and Lennard-Jones terms, and a harmonic intramolecular potential. 
The advantage is that this is well suited for test
purposes, as {\it e.g.} results for the water dimer already exist and can thus
be compared with.
The {\it ab-initio} calculations were done on the simple LDA 
(local density approximation) level, with Dirac-Slater exchange and
the correlation functional as in Ref. \cite{PZ81}. 
If not stated otherwise,
for oxygen a $[4s3p]$ Gaussian basis set (as in Ref. \cite{Towler1994}),
and for hydrogen a $[2s1p]$
basis set (as in Ref. \cite{Ditchfield1971})
was used. These basis sets are medium sized, which
also means that they are robust and problems from linear dependencies
are unlikely to occur.

The structures are
visualised with XCrysDen \cite{XCrysDen} and Molden \cite{Molden}.

\section{Results and discussion}

\subsection{Structural optimisation}

\label{optimisationsection}

\subsubsection{Water molecule and dimer} 

\label{Water_molecule_and_dimer}

Results for the geometry and vibrational frequencies of the
water molecule are displayed in table
\ref{H2Omoleculegeometry}. The MM results are practically identical to those of
the publication where this potential was suggested \cite{ToukanRahman1985}.
The QM results (on the LDA level) are in reasonable agreement with
the LDA results in Ref. \cite{Xu2004}. Larger basis sets were used in Ref. 
\cite{Xu2004},
and indeed, the agreement of the computed H-O-H angle
(108.1$^\circ$ vs 104.9$^\circ$ in Ref. \cite{Xu2004}) can be improved, when
the oxygen basis set is enhanced. For this purpose, the $[4s3p2d]$ oxygen basis
set from Ref. \cite{Valenzano2006} was used, together with a finer
integration grid and higher thresholds for the integral selection.
The O-H distance agrees already well
with the medium sized basis sets described in section \ref{technicaldetails}. 
With the enhanced basis set, also the vibrational frequencies agree
well with the ones computed in Ref. \cite{Xu2004}. 

\begin{table}
\begin{center}
\caption{\label{H2Omoleculegeometry}Geometry of the H$_2$O molecule,
vibrational frequencies and charges at the MM level calculated
with the empirical potential and with QM on LDA-level. For the MM
system, charges are as defined by the potential, while Mulliken charges
are given for the QM case.}
\vspace{5mm}
\begin{tabular}{ccccc}
\hline\hline
\multicolumn{1}{c}{property} & {MM} & QM & QM (enhanced
basis set) & Ref. \cite{Xu2004} (LDA) \\ \hline
distance  O-H1 (O-H2 is equal) 
& 1.000 \AA & 0.971 \AA & 0.971 \AA & 0.970 \AA \\
distance H1-H2 & 1.633 \AA & 1.572 \AA & 1.542 \AA &  \\ \hline
angle H1-O-H2 & 109.5$^\circ$ & 108.1$^\circ$ & 105.1$^\circ$ & 104.9$^\circ$
\\
\hline
frequencies & 1653 & 1436 & 1566 & 1560 \\
in cm$^{-1}$ & 3822 & 3675 & 3712 & 3729 \\
           & 3952 & 3837 & 3826 & 3835 \\ \hline
charge O  &  -0.82 $|e|$ & -0.60 $|e|$ & -0.65 $|e|$ \\
charge H1,H2 & 0.41 $|e|$ & 0.30 $|e|$ & 0.33 $|e|$ \\ \hline
total energy, in $E_h$ & 0 & -75.8763 & -75.8883 &  \\
\hline\hline
\end{tabular}
\end{center}
\end{table}

There are many studies on the equilibrium geometry of the water
dimer, {\it e.g.} based on M{\o}ller-Plesset perturbation theory 
\cite{Smith1990,Schuetz1997}, 
with a local correlation treatment \cite{Schuetz1998},
coupled cluster methods \cite{Klopper2000,Tschumper2002}
or density functional theory \cite{Anderson2006}.
Numerous model potentials have been developed, see for example Ref.
\cite{Jorgensen1981}.
For results on the water dimer with the model used in the present work,
see Ref.  \cite{Kiss2009}, and for an overview of results with model potentials,
see Ref. \cite{Yu2004}.

The optimisation at the MM \cite{Gale1997} and QM 
\cite{IJQC,CPC,Mimmo2001} level is routine and can be done with
the codes used. The QM/MM level is more challenging due to the lack
of the QM Hessian, and thus
various possibilities for the optimisation of a (H$_2$O)$_2$ dimer were 
tested. Most successful turned out to be 
the Broyden-Fletcher-Goldfarb-Shanno (BFGS) or 
Davidon-Fletcher-Powell (DFP) optimiser in combination
with the force minimisation option, 
which means that the gradient norm
is the quantity to be minimised, instead of the energy.
The conjugate gradient minimisation in combination with force minimisation
also worked, but needed many more iterations.

\begin{figure}
\includegraphics[width=9cm,angle=90]{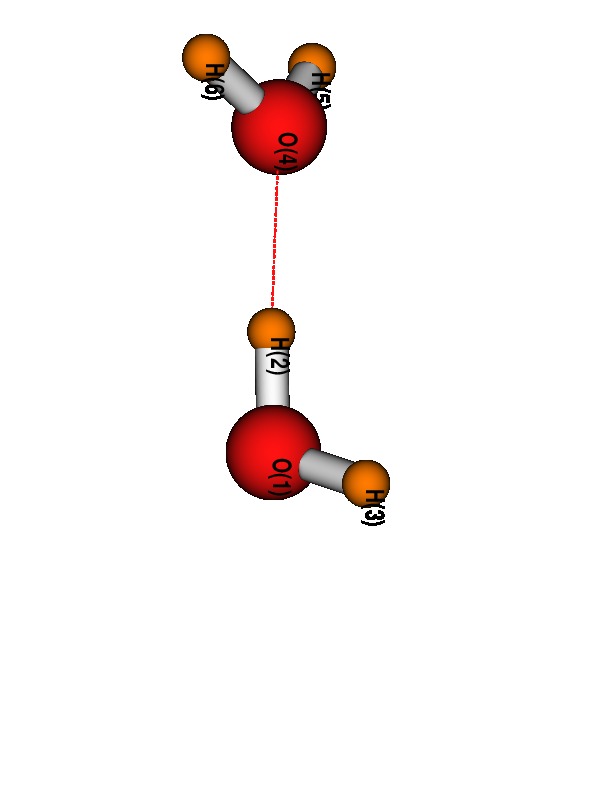}
\caption{ (Colour online) 
MM equilibrium geometry of the water dimer.
Oxygen is displayed as a big red sphere.
The QM/MM equilibrium geometry is
similar. The hydrogen atom H(2) is in the latter two cases
below ('below' according to the arrangement in the figure) the O-O-axis, whereas
in the case of the QM geometry, the hydrogen is above the O-O axis.}
\label{H2Odimer}
\end{figure}

Structural parameters for the (H$_2$O)$_2$ dimer are displayed in 
table \ref{H2Odimergeometry}.
In this dimer, as displayed in figure \ref{H2Odimer},
the right H$_2$O is considered as the QM region.

\begin{table}
\begin{center}
\caption{\label{H2Odimergeometry}Geometry of the (H$_2$O)$_2$ dimer,
with the empirical potential (MM), with QM at the LDA-level
and at the QM/MM level.
The labels of the atoms are as in figure \ref{H2Odimer}, and O1, H2, H3
is the QM region.}
\vspace{5mm}
\begin{tabular}{cccc}
\hline\hline
\multicolumn{1}{c}{property} &
\multicolumn{1}{c}{MM} & {QM} & {QM/MM}  \\ \hline
distance O1-H2 & 1.017 \AA & 0.993 \AA&  0.990 \AA \\
distance O1-H3 & 0.999 \AA & 0.970 \AA & 0.971 \AA  \\
distance O1-O4 & 2.738 \AA & 2.651 \AA  & 2.753 \AA  \\ 
distance O4-H5 (O4-H6 is equal) & 1.003 \AA & 0.972 \AA & 1.003 \AA  \\
angle H2-O1-H3 & 107.6$^\circ$ & 108.8$^\circ$ & 105.2$^\circ$  \\
angle H5-O4-H6 & 108.5$^\circ$ & 109.7$^\circ$ & 108.5$^\circ$  \\ 
angle O1-H2-O4 & 176.1$^\circ$ & (360-174.9)$^\circ$ & 175.1$^\circ$  \\ \hline
total energy, in $E_h$ & -0.0109 & -151.7732  & -75.8866 \\
\hline\hline
\end{tabular}
\end{center}
\end{table}

The MM results for the dimer are in full agreement with an earlier work
where the same potential was used \cite{Kiss2009}, which can be seen
as a test of the implementation of the potential
(the computed MM oxygen-oxygen equilibrium
distance in \cite{Kiss2009} is 2.737 \AA, and the computed MM intermolecular
energy is 28.63 kJ/mol, corresponding to 0.0109 $E_h$ or 0.30 eV).
The QM results show
that there is a stronger attraction on this level, resulting in a
shorter oxygen-oxygen equilibrium bond length. The results agree reasonably
well with the LDA results from \cite{Xu2004}, where an oxygen-oxygen
equilibrium
distance of 2.710 \AA \ was obtained and a binding energy of 9.02 kcal/mol,
corresponding to 0.39 eV or 0.0144 $E_h$. 
Using data from tables \ref{H2Odimergeometry} and
\ref{H2Omoleculegeometry} to compute the binding energy,
then the binding energy would be 0.0206 $E_h$ or 0.56 eV; the deviations are
again due to the smaller basis sets used in the present work. The MM
value (0.0109 $E_h$, 0.30 eV, see above) 
is close to the experiment \cite{Curtiss1979}, whereas LDA overbinds.
The QM/MM results are very
similar to the MM results for the molecule made of O4, H5, H6, {\it i.e.}
the molecule which is not in the QM region. This is to be expected, as
the QM part mainly influences the molecule in the QM region, and the
intermolecular interactions are due to the MM part. 
The QM part has thus only an indirect influence due to the slightly different
geometry of the QM water molecule. Thus, the oxygen-oxygen
equilibrium distance obtained with the QM/MM approach is close to the MM value.

\begin{figure}
\includegraphics[width=12cm,angle=0]{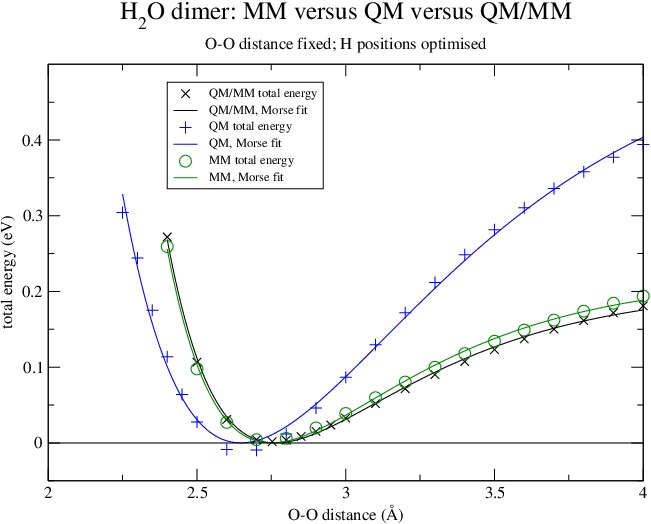}
\caption{ (Colour online) 
The total energy
for a water dimer, on the level of MM, QM and QM/MM.
All hydrogen positions are optimised for a fixed oxygen distance, and 
the potential curve is obtained as a function of the oxygen-oxygen distance.
The minimum energy of the various fits is shifted to zero.}
\label{H2Odimerpotentialcurve}
\end{figure}

Subsequently, the energy curve when separating both water molecules
was computed. For this purpose,
the oxygen-oxygen distance was fixed at a certain value, and the geometries of
all hydrogen positions were optimised. This way, one point of the potential
curve is obtained. When this procedure is performed for a set of 
oxygen-oxygen distances, then the potential curve in figure
\ref{H2Odimerpotentialcurve} is obtained. There, the data was fitted
with a Morse potential. It turns out that MM and QM/MM are very similar
(equilibrium distance, curvature at the minimum, binding energy of the dimer).
This is due to the reason mentioned above, that the QM part has only little
influence on the intermolecular interaction.
From the fits with the Morse potential,
the binding energy is obtained as
0.22 eV (MM), 0.21 eV (QM/MM) versus 0.55 eV (QM).

\subsubsection{Two dimensional periodicity}

In a next step, periodicity was implemented. 
An advantage of the presently used codes is that both
employ the same electrostatic potentials,
{\it i.e.} Saunders' potential \cite{Vic1994} 
for one dimensional periodicity, Parry's \cite{Parry} for
two dimensional periodicity and the Ewald potential for three 
dimensional periodicity. Note that in the present
case of two-dimensional periodicity, the implementation of the
periodicity is truly two-dimensional. This implies that
the system is {\it not} repeated in the third direction ({\it z}-direction),
and there is consequently no parameter for the vacuum thickness.

\begin{figure}[h]
\includegraphics[width=12cm,angle=0]{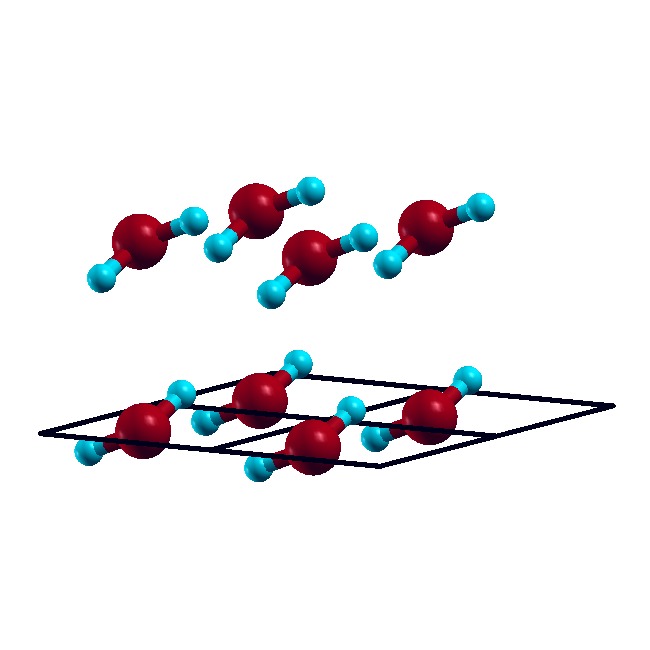}
\caption{ (Colour online) 
A periodic arrangement of a water dimer, with lattice constant {\it a=b}
(in the present figure: {\it a=b}=3 \AA) and
a fixed angle $\gamma=90^\circ$. Note that the system is periodic only in
two dimensions, but not in the {\it z}-direction.}
\label{H2Odimer2dperiodicgeometry}
\end{figure}

\begin{figure}[h]
\includegraphics[width=12cm,angle=0]{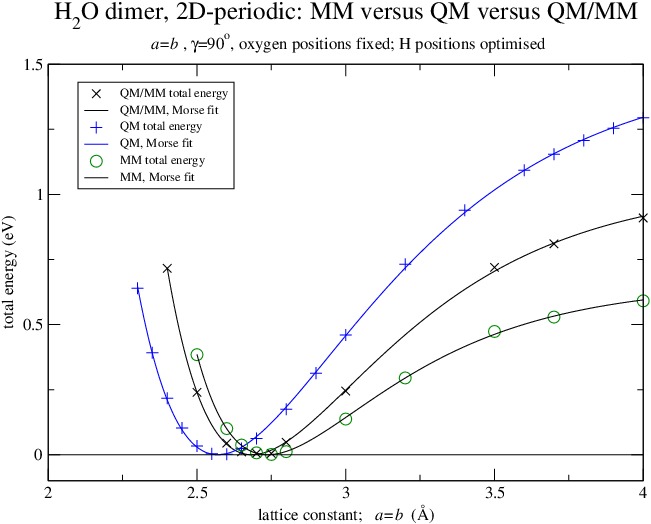}
\caption{ (Colour online) 
The total energy
for a water dimer, arranged with two-dimensional
periodicity, on the level of MM, QM and QM/MM.
All hydrogen positions are optimised for fixed oxygen positions, and 
the potential curve is obtained as a function of the lattice constant,
with the lattice constant {\it a=b} and a fixed angle $\gamma=90^\circ$.
The minimum energy of the various fits is shifted to zero.}
\label{H2Odimer2dperiodicpotentialcurve}
\end{figure}

As a test, the potential curve was computed,
where a water dimer was periodically arranged, with the lattice constants
$a=b$, and a fixed angle $\gamma=90^\circ$ 
(see figure \ref{H2Odimer2dperiodicgeometry}). 
Then, the oxygen positions were
fixed at a distance of 2.8 \AA \ (one oxygen 2.8 \AA \ vertically above the
other along the {\it z}-axis, which is the non-periodic direction).
For various
lattice constants, the hydrogen positions were optimised. In the QM/MM
case, one molecule was considered the QM region; here it is the one
whose oxygen atom has a hydrogen bridge to the second water 
molecule. This resulted in
three potential curves visualised in figure 
\ref{H2Odimer2dperiodicpotentialcurve}.

For large values of the lattice constant, one obtains essentially the
structure of a single water dimer, as the interaction between the dimers
is small. When reducing the lattice constant, the water molecules tilt
to the molecules of the neighbouring cell, as the interaction to this
neighbouring molecule becomes more important.
The potential curve has a minimum for a certain lattice constant, which
is shortest at the QM level and longer at the MM level. The
QM/MM equilibrium distance is now in between the QM and MM equilibrium
distance, as the dominant interactions are now the
QM-interaction between neighbouring molecules in one layer and 
the MM-interaction between neighbouring molecules in the other layer.

\subsection{Molecular dynamics}

In a next step, the QM/MM expressions as in equations
(\ref{energyequation}) and (\ref{gradientequation})
were used to perform molecular dynamics. The implementation is similar
to the one used for geometry optimisation as was described in the previous
sections. The energy and gradient calculations
in the molecular dynamics part of the GULP code were correspondingly
modified, and the {\it ab-initio} energy and gradient are again extracted
from CRYSTAL. 

\subsubsection{Water dimer}

To test the implementation, molecular dynamics simulations were performed for
the water dimer (without any periodicity), 
at the level of MM, QM and QM/MM. The 
starting point was a system consisting of
two water molecules at a relatively large distance
(oxygen-oxygen distance $\sim$5 \AA).
At the MM level, as CPU time is not critical, a 
simple, nearly brute force way to obtain the minimum structure was
to employ an NVT ensemble at $T$=10 K 
with a production run of 400 ps, using a time step of 1 fs. 

On the QM level, the optimisation was more
difficult due to problems with overheating, when the system gained
kinetic energy too quickly. Thus,   
a series of molecular dynamics runs was used to obtain the minimum structure. 
Starting again from the
geometry where two water molecules are far apart,
a 1 ps simulation was performed (time step 1 fs, initial
temperature 100 K, and cooling with a rate of 0.1 K/step down to 5 K).
Subsequently, the final geometry was used as a new initial geometry,
and a simulation at a temperature of 5 K (NVE ensemble, 1 ps, time step 1 fs)
was performed. After repeating this procedure a few times, the final
structure was essentially the one of the geometry optimisation in 
table \ref{H2Omoleculegeometry}. 

At the QM/MM level, the simulation turned out to be straightforward. 
The NVT ensemble
was used, at a temperature of 10 K, and with a simulation length of 100 ps
(1 fs time step). 

In all cases, the final structure agreed reasonably well with the
one obtained from a straightforward optimisation as described in section
\ref{optimisationsection}, {\it e.g.} the O-O distance at the end of the
MD run was 2.733 \AA \ (MM), 2.614 \AA \ (QM), and 2.781 \AA \ (QM/MM).

\subsubsection{Two-dimensional periodicity}

Two water molecules were periodically arranged in a two-dimensional
box of 5$\times$5 \AA \ size, and a molecular dynamics
run at the MM level
was performed (NVT ensemble, 400 ps total length, 1 fs time step). 
The initial temperature was 300 K, and lowered to 50 K in 25000 steps.
The final
structure agreed with the one obtained by a normal minimisation with the default
optimiser (BFGS). With a similar (though shorter) 
molecular dynamics run and using
the QM/MM energy and forces, an optimisation at the QM/MM level was
performed. The structure agreed reasonably well with the one from the
MM optimisation. This serves as an additional test for the QM/MM implementation.

\subsubsection{Three-dimensional periodicity}

\begin{figure}[h]
\includegraphics[width=11cm,angle=0]
{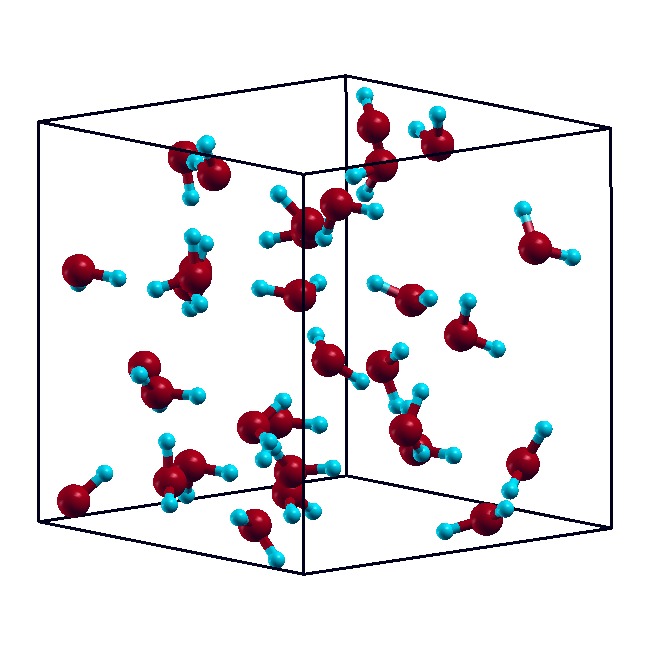}
\caption{ (Colour online) A snapshot of the molecular dynamics simulation of
a box containing 32 water molecules.}
\vspace{1.0cm} 
\label{waterbox}
\end{figure}

In the case of three-dimensional periodicity, 32 water molecules were
considered, in a cubic box of 9.856 \AA, corresponding to a density
of 1 g/cm$^3$ (see figure \ref{waterbox}). The calculation of the
radial pair distribution function and of the mean square displacement (MSD)
was implemented in GULP, or independently computed with ISAACS
\cite{LeRoux2010}. A simulation length of 25 ps equilibration (NVT
ensemble) and subsequently
100 ps production was chosen (NVE ensemble), at a time step of 1 fs.

\begin{figure}[h]
\includegraphics[width=10cm,angle=270]
{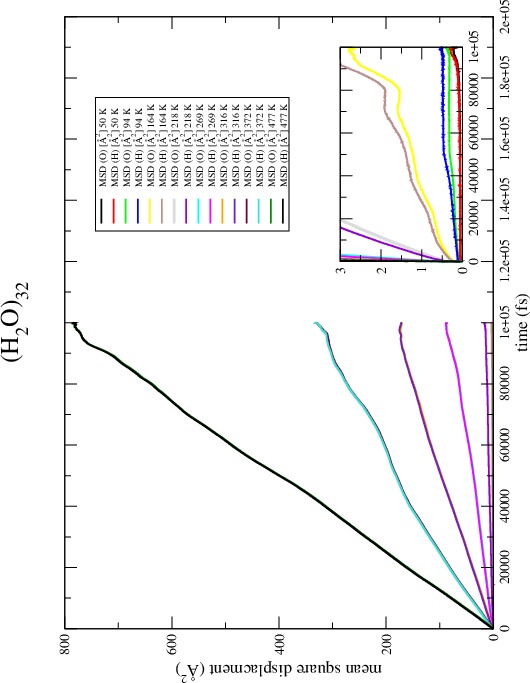}
\caption{ (Colour online) The mean square displacement (MSD) for a system of
32 H$_2$O molecules, at the MM level, for temperatures in the range
from 50 to 491 K. The MSD for temperatures $\lesssim$ 164 K is almost constant
and displayed in the inset.}
\label{MMmsd}
\end{figure}

\newpage

\begin{figure}

a) \hspace{1cm} \includegraphics[width=6.3cm,angle=270]
{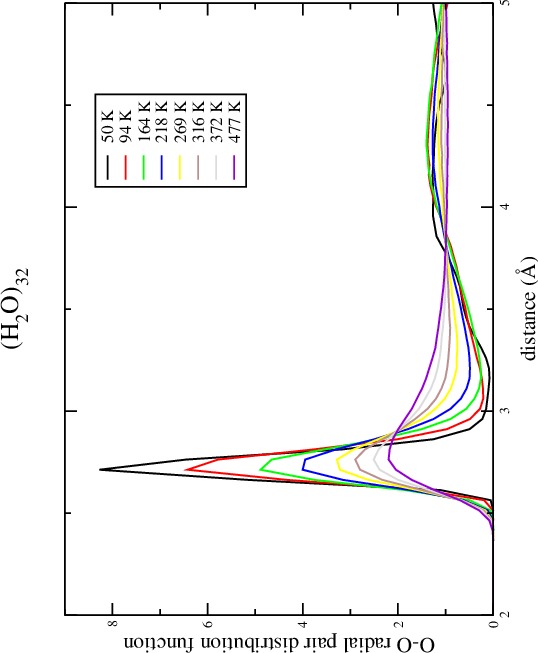}

b) \hspace{1cm} \includegraphics[width=6.3cm,angle=270]
{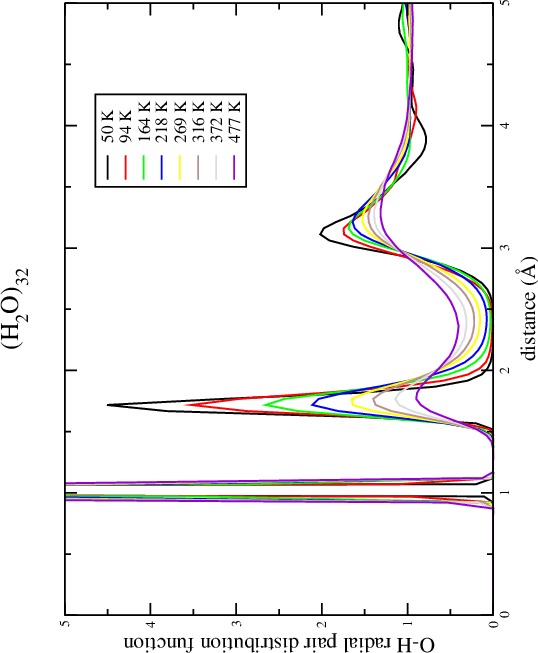}

c) \hspace{1cm} \includegraphics[width=6.3cm,angle=270]
{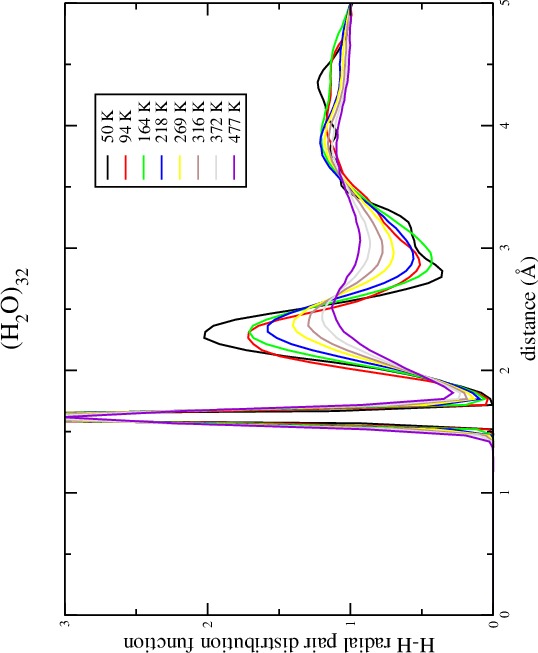}
\caption{ (Colour online) The radial pair distribution function
for a system of
32 H$_2$O molecules, at the MM level, for temperatures in the range
from 50 to 477 K. The displayed radial pair distribution functions
are for O-O (a), O-H (b) and H-H (c).}
\label{radialPDFbild}
\end{figure}

Results for the MSD at the MM level are summarised in figure \ref{MMmsd}.
The temperature is the one as it was obtained as an average over the run.
For temperatures $ \gtrsim$ 218 K, there is a linear slope, which
increases with temperature, whereas for temperatures $\lesssim$ 164 K,
the MSD is approximately constant. This indicates a phase transition
from the liquid to the solid phase in this temperature range, with this
potential. 
The radial pair distribution functions are displayed in figure
\ref{radialPDFbild}. They agree reasonably well with the literature,
{\it e.g.} with empirical potentials \cite{Rahman1973,Jorgensen1983,Sprik1991},
DFT \cite{Laaseonen1993,Izvekov2002,Vandevoldele2005} and recently
also MP2 \cite{DelBen2013}.

\begin{figure}[h]
\includegraphics[width=12cm,angle=0]
{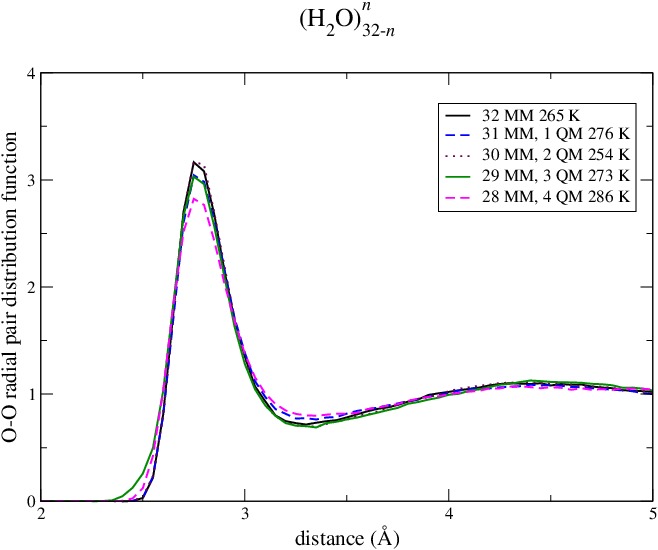}
\caption{ (Colour online) The radial pair distribution function
for a system of
32 H$_2$O molecules, at the QM/MM level, with the number of QM molecules
varied from 0 to 4.}
\label{radialPDFbildQMMM}
\end{figure}

Finally, QM/MM simulations were performed for systems with 32 H$_2$O
molecules, out of which 0 to 4 molecules were treated on the QM level, and all
the other ones on the MM level. This is labelled as
(H$_2$O)$^n_{32-n}$ (where superscript and subscript are the numbers
of QM and MM treated molecules, respectively), with $n$ ranging
from 0 to 4. The QM
region is
fixed to the initially defined QM molecules. More sophisticated approaches
define a spatial QM region, and consider diffusion into and out of this
region via an adaptive scheme \cite{Bulo2009,Bulo2013}.
The present results are similar to
the MM results, for example a comparison is shown in figure 
\ref{radialPDFbildQMMM} for the oxygen-oxygen radial pair
distribution function. 
In the simulations,
it turns out that QM clusters are formed: initially, the QM molecules
were put far apart. After a while, they start to form dimers in the
case of (H$_2$O)$_{30}^2$, and arrange in a triangle for 
(H$_2$O)$_{29}^3$. For (H$_2$O)$_{28}^4$, there are enough QM
molecules so that they can form a chain, which connects the cells; {\it i.e.}
there is a stripe of QM molecules going through the unit cell. This 
formation of QM clusters and QM stripes is presumably due
to the different binding energies of QM molecules and MM molecules
with the functional and potential used in the present work (see the
discussion about the water dimer in section \ref{Water_molecule_and_dimer}).
This may require a readjustment of the water potential is some cases, but not
necessarily in general. For example, in the case of
 systems such as water in contact
with a transition metal surface, the first water layer has an 
ice-like structure, and might constitute 
a possible QM region, whereas the remaining
water molecules might be described with a semi-empirical potential.

\section{Conclusion}
A QM/MM implementation for periodic systems has been presented.
Calculations on the water dimer and for water with two and three-dimensional
periodicity show that the implementation works properly. 
Structural optimisation
and molecular dynamics simulations have been implemented on the QM/MM level.
In the case of three-dimensional
periodicity, as a test system
a box of 32 water molecules was studied at the MM and QM/MM
level. The pair distribution functions
agree well when computed on the MM or QM/MM levels. 

\section*{Acknowledgements}

We acknowledge support from the European Research Council (ERC) through the
ERC-Starting Grant THEOFUN (Grant Agreement No. 259608) as well as from the
DFG (Deutsche Forschungsgemeinschaft).

\clearpage

\end{document}